# Gravitational waves and core-collapse supernovae

G.S. Bisnovatyi-Kogan(1,2), S.G. Moiseenko(1)
(1)Space Research Institute, Profsoyznaya str/ 84/32, Moscow 117997, Russia
(2)National Research Nuclear University MEPhI, Kashirskoe shosse 32,115409 Moscow, Russia

**Abstract.** A mechanism of formation of gravitational waves in the Universe is considered for a nonspherical collapse of matter. Nonspherical collapse results are presented for a uniform spheroid of dust and a finite-entropy spheroid. Numerical simulation results on core-collapse supernova explosions are presented for the neutrino and magneto-rotational models. These results are used to estimate the dimensionless amplitude of the gravitational wave with a frequency $v \sim 1300$ Hz, radiated during the collapse of the rotating core of a pre-supernova with a mass of $1.2\,M_\odot$ (calculated by the authors in 2D). This estimate agrees well with many other calculations (presented in this paper) that have been done in 2D and 3D settings and which rely on more exact and sophisticated calculations of the gravitational wave amplitude. The formation of the large-scale structure of the Universe in the Zel'dovich pancake model involves the emission of very long-wavelength gravitational waves. The average amplitude of these waves is calculated from the simulation, in the uniform spheroid approximation, of the nonspherical collapse of noncollisional dust matter, which imitates dark matter. It is noted that a gravitational wave radiated during a core-collapse supernova explosion in our Galaxy has a sufficient amplitude to be detected by existing gravitational wave telescopes.

## 1. Introduction

On February 11, 2016, LIGO (Laser Interferometric Gravitational-wave Observatory) in the USA with great fanfare announced the registration of a gravitational wave (GW) signal on September 14, 2015 [1]. A fantastic coincidence is that the discovery was made exactly 100 years after the prediction of GWs by Albert Einstein on the basis of his theory of General Relativity (GR) (the theory of space and time). A detailed discussion of the results of this experiment and related problems can be found in [2–6].

Gravitational waves can be emitted by binary stars due to their relative motion or by collapsing nonspherical bodies. The GW signal is very difficult to detect due to the extreme weakness of gravitational interaction. Gravitation on Earth can be detected because it is produced by an enormous mass of matter and cannot be screened. If we hypothetically imagine Earth without electrons, made of protons only, the electric attraction (or repulsion) force of protons would be by 36 orders of magnitude higher than the gravitational attraction force. In fact, due to the presence of electrons, Earth is electrically neutral, and the corresponding electric force does not exceed the gravitational force.

When registering a GW signal, a tiny displacement of masses was measured, corresponding to the change in a meter-long body by an amount about 10 orders of magnitude smaller than the hydrogen atom size. Different perturbations produced by surrounding bodies or terrestrial seismic noises cause displacements in the detector that can significantly exceed the GW signal. To minimize background noises of local origin, several (at least two) detectors separated by a large distance are used.

The LIGO observatory includes two laser detectors located in Louisiana and Washington states in the USA at a distance of 3000 km from each other. The extraction of the signal from the background is performed by correlating signals on both detectors. A GW signal should be totally identical on two detectors because it propagates almost without distortion from the source to Earth. Because the GW propagation speed should be equal to the speed of light, there is a time delay between the times of arrival of the signal at the two detectors. The observed time delay of about 10 ms enabled identifying a strip on the sky with the area of about 600 square degrees from which the signal could arrive. Such a big area in the sky for the signal greatly complicates its identification with some visible source in the sky.

A comparison of the signal waveform detected in both detectors with theoretical calculations allows estimating the masses of the coalescing objects emitting the GW signal. These objects turned out to be two black holes with masses of 36 and 29 solar masses ($M_\odot$). The theory allows estimating the gravitational wave power radiated during the coalescence of such black holes, and by comparing it with the registered signal it is possible to estimate the distance to the source, about 400 Mpc, corresponding to the redshift $z = 0.09$. The energy carried out by the GWs is enormous: it is equivalent to the rest-mass energy of a $3M_\odot$ body. The GW was registered by the advanced LIGO detectors, which are able to detect a signal producing a relative deformation $h \sim 10^{-23}$. The detected signal turned out to be much stronger than this threshold, and it might have been detected by the same installation before its upgrade and by some other detectors. By chance, none of these detectors (in Italy, Japan, and Germany) was operating at this time.

A system of two massive black holes has never been observed before and has not been discussed much as a possible GW source (see, however, [7, 8]). The most secure GW sources were thought to be binary neutron stars with orbital periods of a few hours, which merge due to the GW emission and produce a powerful GW pulse prior to the coalescence. Such systems are observed in our Galaxy as binary radio pulsars. Calculations of their lifetimes before the coalescence, as well as statistics on such sources in other galaxies, gave estimates (by an order of magnitude) of their detection rate in the surrounding $\sim 200$ Mpc volume of the Universe of about ten events per year (at the registration threshold level) [9, 10] (see also [11]). The detected GW signal exceeded the threshold 24-fold, although it came from a distance of 400 Mpc. That is, the signal turned out to be about 1000 times as powerful as expected from a binary neutron star coalescence.

The advanced LIGO signal may be the first laboratory detection of GWs. However, indirect detection of GWs following from the analysis of radio observation of a binary Hulse–Taylor pulsar discovered in 1975 [12] was made as early as the end of the 1980s [13] and later confirmed by observations of a close binary system of two radio pulsars discovered in 2004 [14]. The decrease in the orbital period of this system due to GW emission coincides to within $\sim 0.01\%$ (the experimental accuracy) with GR predictions.

Interestingly, the first registration of GW signals was announced by American physicist Joseph Weber in 1969 [15]. Weber constructed two solid-body cylinders capable of registering oscillations, which were separated by 1000 km from each other, and searched for a correlated signal. Weber claimed a measurement of relative displacements at the level of $10^{-16}$. This was 100 thousand times as high as the displacement measured by LIGO. Such a strong GW signal contradicted all existing theories and facts. Although Weber's result was experimentally refuted several years after, it induced a whole wave of gravitational experiments, on the crest of which the epochal discovery was made.

The discovery of GWs on Earth resulted from solving an extremely complicated technical and technological problem. The scientific significance of this discovery from the standpoint of fundamental science is not very high, because the validity of GR, underlying the existence of GWs, and their indirect detection in binary systems with radio pulsars have been observationally established with a maximum possible accuracy in radio astronomical experiments.

In this paper, we report on the results of studies of GW generation during nonspherical stellar collapses accompanied by supernova explosions, as well as during nonspherical collapse of large masses associated with the formation of the large-scale structure of the Universe. The results of different calculations showed that GWs generated in a galactic core-collapse supernova explosion can be registered by the LIGO and Virgo detectors.

## 2. Gravitational wave emission during nonspherical collapse

The gravitational energy emitted during a nonspherical collapse was estimated in [16], where the collapse of a homogeneous rigidly rotating dust cloud was considered. The GW power from an oblate spheroid with a mass $M$, a major semi-axis $A$, and a minor semi-axis $C$ can be written as [17, 18]

$$L_{\mathrm{GW}} = \frac{2}{375} \frac{GM^2}{c^2} (\dddot{A}^2 - \dddot{C}^2)^2. \tag{1}$$

In the case of collapse of a nonrotating body, the emitted GW energy is given in [16] as

$$E_{\mathrm{GW}} = 0.0370 \left( \frac{r_{\mathrm{g}}}{A_{\min}} \right)^{7/2} Mc^2 \lesssim 10^{51} \text{ erg}. \tag{2}$$

Here, $r_{\mathrm{g}} = 2GM/c^2$ is the Schwarzschild radius and $A_{\min}$ is the minimal value of the major semi-axis. In the case of rapid rotation, less energy is emitted because the rotation leads to a bounce (at $C = 0$) with a lower surface density and higher value of $A_{\min}$:

$$E_{\mathrm{GW}} = 0.109 \left( \frac{r_{\mathrm{g}}}{A_{\min}} \right)^{7/2} Mc^2 \gtrsim 10^{45} \text{ erg}. \tag{3}$$

It is noted in [19] that most of the GW energy is emitted during the matter bounce. To avoid an infinite density at the moment of bounce during the dust cloud collapse, a finite-entropy spheroid was considered, during the collapse of which the density at the bounce remains finite and the thickness is nonzero. As a result, the following formula was obtained [19]:

$$E_{\mathrm{GW}} \approx k M c^2 \left( \frac{r_{\mathrm{g}}}{A_{\min}} \right)^{7/2} \frac{A_{\min}}{C_{\min}}. \tag{4}$$

Here, according to estimates, $k \sim 0.01$ [19], and $C_{\min}$ is the minimal value of the minor semi-axis. Formula (4) leads to a nonphysical infinity as $C_{\min} \to 0$ with $A_{\min}/C_{\min} \to \infty$. The estimate of the minimal value of $C_{\min}$ at which formula (4) is valid can be obtained by comparing (4) with formula (2) for a nonrotating zero-energy dust spheroid, to which the emitted GW energy should approximately tend as $C_{\min} \to 0$ due to the vanishing of entropy, and with formula (1) for a rapidly rotating spheroid. As a result, we obtain the respective limits for nonrotating and rapidly rotating spheroids:

$$\frac{A_{\min}}{C_{\min}} \lesssim 4, \quad \frac{A_{\min}}{C_{\min}} \lesssim 11. \tag{5}$$

A detailed study of gravitational radiation from a collapsing spheroid is performed in [20].

## 3. Core-collapse supernova models without magnetic fields

Core-collapse supernovae are possible GW sources. The collapse of the pre-supernova iron core occurs nonsymmetrically. Presently, the physical mechanism of supernova explosions is not known in detail. At the late stages of the evolution of massive stars, the loss of hydrodynamic stability initiates the collapse enabling the supernova explosion and neutron star formation. An enormous amount of energy is released, equal to the bound energy of a neutron star, $(0.15-0.2)\,Mc^2$. Most of this energy is carried away by weakly interacting neutrinos, which escape freely. A small part of the energy of the neutrino flux is absorbed by and heats the surrounding envelope, which can lead to the formation of a shock leading to the supernova explosion. The neutrino model of supernova explosions was first calculated by Colgate and White [21] and has been considered many times in a one-dimensional setup in papers [22–26] and many others. No explosion has been obtained in the one-dimensional spherically symmetric model; therefore, in recent years research has been focused on two- and three-dimensional models.

Hopes to obtain explosion in multi-dimensional neutrino models are based on the convective instability of the collapsing core first discovered in [27]. Convection carries more energetic neutrinos away from the inner parts of the core. Because the neutrino interaction cross section increases with energy as $\sim E_\nu^2$, the heating of the envelope increases and the probability of explosion becomes higher. We note that this effect depends on the size of convective eddies: the higher they are, the higher the energy of neutrinos carried away and the higher the explosion efficiency. For quantitative estimates, two- and three-dimensional calculations are required. The results of various authors differ greatly. The results of core-collapse and supernova explosion calculations using two-dimensional hydrodynamics (piecewise parabolic method, PPM) and two-dimensional neutrino transfer are presented in [28]. In that paper, a strong convection behind the bounced shock front was obtained, which gave sufficient power for a strongly asymmetric explosion. The two-dimensional modeling of the collapse and supernova explosion using the PPM carried out in other similar studies [29, 30] produced no powerful explosion. Two-dimensional supernova explosion calculations using the smooth particle hydrodynamics (SPH) method were performed in [31]. In that paper, a supernova explosion was obtained; however, calculations carried out in [32] under the same setup but using a different numerical scheme showed that no explosion can be produced.

In the three-dimensional setup, the supernova explosion problem was modeled by the SPH method [33]. According to that paper, as in an earlier paper [31], a supernova explosion occurs. When modeling supernovae by the SPH method, the explosion occurs at the stage where the applicability of this numerical method is unjustified, because the atmosphere surrounding the forming neutron star contains too few particles, and the resolution of the method is low.

Two-dimensional calculations with physical effects taken into account with greater precision are presented in [34], where rotation and neutrino convection are taken into account, and neutrino losses are calculated for the first time from the solution of the Boltzmann equation. The results of these calculations showed that core collapse does not lead to a supernova explosion. The emerging shock moves a distance of about 100–200 km from the stellar center and stalls. Presently, neutrino-convection-based numerical calculations of core collapse and supernova explosion highly depend on the numerical method and the setup details, as wells as on the neutrino emission treatment used.

The increasing computer power enabled modeling neutrino-based supernova explosions with a higher spatial resolution. However, in the three-dimensional case, the turbulence leads to spatial fragmentation of turbulent eddies (whereas in the two-dimensional case, the eddies become larger), which decreases the supernova explosion efficiency. Neutrino-based core-collapse supernova explosions are modeled by many groups (see, e.g., [35–41] and the references therein).

In [42], a supernova explosion mechanism is proposed based on the fission of the collapsing core into two parts, with one part being a neutron star. Due to gravitational radiation, the parts of the separated core approach until the less massive piece fills its Roche lobe. Then mass transfer begins. When the mass of the less massive component decreases to the lower mass limit for neutron stars, an explosive deneutronization of the low-mass neutron star can occur. This energy release can expel the envelope of the collapsing star. For this mechanism to operate, a very rapid pre-supernova rotation is required. Presently, the direct numerical modeling of this supernova explosion mechanism is difficult to perform.

## 4. Magnetorotational supernovae

In 1970, Bisnovatyi-Kogan [43] proposed a magnetorotational mechanism of core-collapse supernova explosions, in which rotation and the magnetic field play crucial roles. Part of the gravitational energy released in the collapse goes into the rotation and magnetic field energy. The differential rotation and frozen magnetic field allow the transformation of this gravitational energy into radial kinetic energy—the energy of explosion. Presently, the magnetorotational explosion model most successfully explains the energy release in core-collapse supernovae that can be accompanied by the formation of directed mass ejections—jets.

To model the magnetorotational supernova explosion, a system of equations of gravitational magnetic hydrodynamics with infinite electric conductivity (ideal MHD) is used. The equation of state includes the contributions from electrons with an arbitrary degree of degeneracy, matter neutronization with a density increase, and strong interactions in the form of nuclear repulsion at densities typical for the central parts of neutron stars. In addition, neutrino emission is taken into account, which freely escapes from the dense core and carries away most of the collapse energy. The first calculations in the one-dimensional cylindrical approximation [44, 45] demonstrated a high efficiency of transformation of the rotational energy into the energy of the explosion mediated by the magnetic field. For two-dimensional calculations, a special code based on an implicit Lagrangian scheme on an adapting triangle mesh was developed [46–48], which was used to perform the first two-dimensional numerical simulations of magnetorotational supernovae [49, 50].

The modeling of a magnetorotational supernova explosion included two consecutive stages. First, the collapse of the iron core with the formation of a differentially rotating neutron star was calculated [51] (Fig. 1). At this stage, the dynamical effect of the magnetic field can be ignored. Then, a poloidal magnetic field was 'turned on', and the calculation of the toroidal magnetic field enhancement during the differential rotation, resulting in the magnetorotational supernova explosion, was computed. The calculations were performed for the initial quadrupole [49] and dipole [50] poloidal magnetic field.

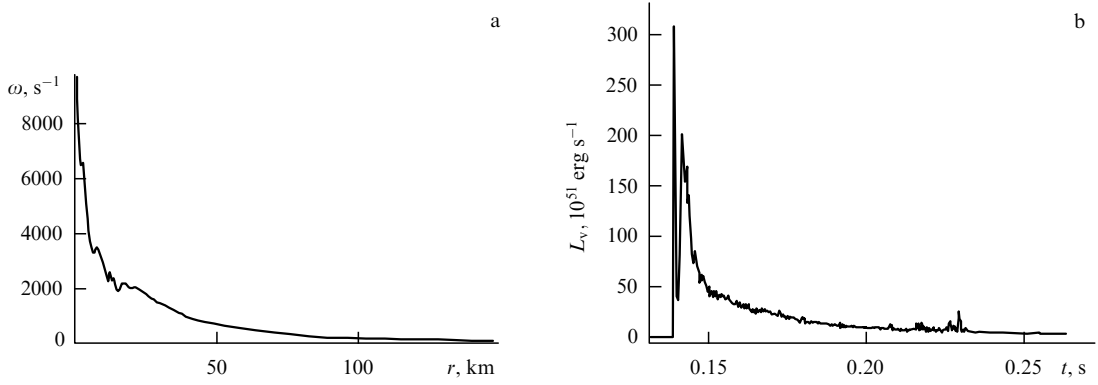

**Figure 1.** (a) Angular velocity of matter $\omega$ of a neutron star as a function of the radius $r$ in the equatorial plane after collapse. (b) Time evolution of the neutrino luminosity from the beginning of the collapse to the formation of a rotating neutron star [51].

At the initial time, an iron $^{56}$Fe core of a pre-supernova star with a mass of $1.2\,M_\odot$, radius of 1370 km, and central density of $4.5 \times 10^9$ g cm$^{-3}$ was considered. Here, the initial rotational-to-gravitational energy ratio and the internal-to-gravitational energy ratio were

$$\frac{E_{\rm rot}}{E_{\rm grav}} = 0.0057, \quad \frac{E_{\rm int}}{E_{\rm grav}} = 0.727\,.$$

At the time $t = 0.1377$ s after the beginning of the collapse, a bounce shock formed at a distance of $6 \times 10^5$ cm from the center. Because of the high temperature behind the shock front, neutrinos were created, which carried away most of the gravitational energy released. Figure 1 shows the neutrino loss rate during the collapse. The density at the stellar center reaches the maximum value $\rho_{\rm c,max} = 5.655 \times 10^{14}$ g cm$^{-3}$ at the moment $t = 0.1424$ s. The shock entrains the stellar matter, ejecting $\sim 0.041\%$ of the core mass and $\sim 0.0012\%$ ($2.960 \times 10^{48}$ erg) of its gravitational energy. As a result, by the time $t = 0.261$ s after the decay of oscillations, a differentially rotating configuration is produced. The central proto-neutron star with a radius $\sim 12.8$ km rotates almost rigidly with a period of 0.00152 s. The angular velocity rapidly decreases with the distance from the center (see Fig. 1), and the rotation period reaches $\sim 35$ s near the outer boundary in the equatorial plane. The mass of the configuration outside the rigid core was $\sim 1\,M$.

The results of the numerical modeling of the core collapse suggest that in agreement with the results of many other authors, the amount of matter ejected during the collapse and the collapse energy carried away are too low to explain the supernova explosion phenomenon. At that moment, the initial poloidal magnetic field was turned on, whose energy was $\sim 10^{-6}$ of the gravitational energy. The magnetic field was taken with either the quadrupole ($H_z = 0$ at $z = 0$) [49] or dipole ($H_r = 0$ at $z = 0$) [50] symmetry.

By means of differential rotation, the radial field velocity in the initial configuration gives rise to a toroidal magnetic field component $H_\varphi$ that linearly increases with time. The toroidal magnetic field energy first increases quadratically with time (Fig. 2). Starting from the time $t = 0.04$ s after turning on the field, both field components start rapidly (exponentially) growing due to differential magnetorotational instability [52]. After the toroidal magnetic field energy reaches the maximum value $4.8 \times 10^{50}$ erg at $t = 0.12$ s, a shock emerges and an explosion occurs due to conversion of the magnetic and rotational energy into radial kinetic energy. The maximum $H_\varphi$ in this process was $\sim 2.5 \times 10^{16}$ G. The poloidal magnetic energy here reached $\approx 2.5 \times 10^{50}$ erg and remained constant nearly at this level until the end of the computation (see Fig. 2).

For the initial quadrupole magnetic field, the MHD shock wave has a large amplitude and moves faster near the equatorial field $z = 0$, leading to a predominantly equatorial mass ejection [49]. Conversely, for the initial dipole field, the main mass ejection occurs along the rotational axis in the form of a mildly collimated jet [50]. Figure 3 shows the dependence of the ejected mass and the kinetic energy on time for the magnetorotational explosion with the quadrupole initial magnetic field. For the dipole field, both these quantities have almost the same value [50]. In both case, the ejection mass is $\sim 0.14\,M_\odot$ and the ejected energy is $\sim 0.6 \times 10^{51}$ erg.

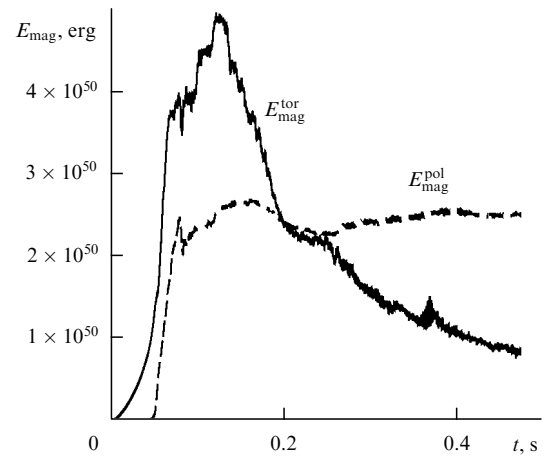

**Figure 2.** Change in the toroidal $E_{\rm mag}^{\rm tor}$ and poloidal $E_{\rm mag}^{\rm pol}$ magnetic field energies during a magnetorotational supernova explosion with the initial quadrupole field [49].

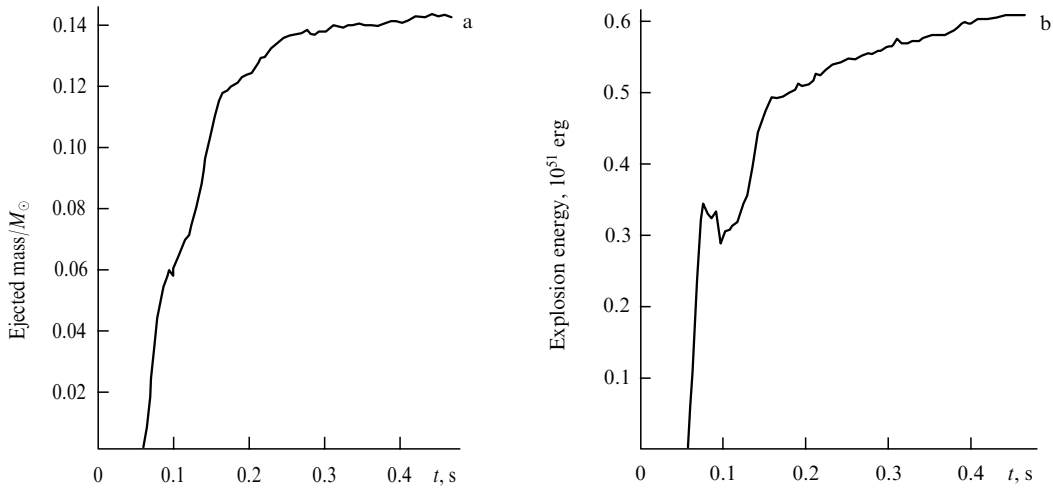

**Figure 3.** (a) The relative mass and (b) energy of ejected matter as a function of time. Both quantities are given for the initial quadrupole field [49].

During an MHD explosion, the star loses a significant part of its rotational energy. The stellar rotational energy not only is transformed into the energy of explosion (the kinetic energy of radial motion), but also is lost as neutrino emission. After the explosion, the core rotates more slowly, and additional compression and some heating of the neutron star occur. Our calculations were stopped at the time $t \approx 1.45$ s, when the proto-neutron star rotated almost rigidly with a period of $\sim 0.006$ s.

Our calculations showed that the amount of energy released during magnetorotational supernova explosions, $0.6 \times 10^{51}$ erg and higher, suffices to explain core-collapse supernova explosions [53]. The young neutron star formed rotates relatively slowly due to the transformation of rotational energy into the explosion energy. Three-dimensional calculations of the magnetorotational explosion were carried out in [54, 55]. The magnetorotational mechanism is based on the action of a magnetic piston, depending on the field strength but not on the scale of turbulent eddies. Thanks to this, the magnetorotational explosion efficiency in two- and three-dimensional calculations are comparable, unlike in the neutrino model.

## 5. Gravitational waves in numerical modeling of core-collapse supernova explosions

Multi-dimensional numerical modeling of core-collapse supernova explosions allowed calculating the amount of gravitational energy emitted and the shape of the GW signal. The GW signal produced in the nonspherical collapse of a rotating body was calculated in [56, 57]. Among the first papers in which the GW signal was calculated from modeling a core-collapse supernova explosion were [58, 59]. To calculate the shape and observable characteristics of the GW signal, the numerical code elaborated in [60] and its modifications were used. In numerical simulations, the quadrupole moment of the collapsing star

$$D_{\alpha\beta} = \int \rho(\mathbf{r}, t)(3r_\alpha r_\beta - \delta_{\alpha\beta} r^2), \qquad (6)$$

and its time derivatives were computed. The GW power $\dot{E}$, which is related to the third time derivative of the quadrupole moment, is [17]

$$\dot{E} = \frac{G}{45 c^5} \, \dddot{D}_{\alpha\beta} \, \dddot{D}^{\alpha\beta} \,. \qquad (7)$$

To calculate the spectrum of GWs, Fourier components of the time-dependent quadrupole moment were used. In [61, 62], GW signal calculations for rotating core-collapse supernovae in two-dimensional numerical simulations are discussed with neutrino processes and the equation of state taken into account in great detail. The GW signal frequency was shown to be weakly dependent on the initial angular momentum and was about 700 Hz for the equation of state specified in [59]. Figure 4 shows the GW amplitude $h$ as a function of time for the distance 10 kpc and the maximum density $\rho_{\max}$ for three models of core-collapse supernovae with different parameters of the initial rotation $\beta_i$:

$$\beta_i = \left| \frac{T}{W} \right| \qquad (8)$$

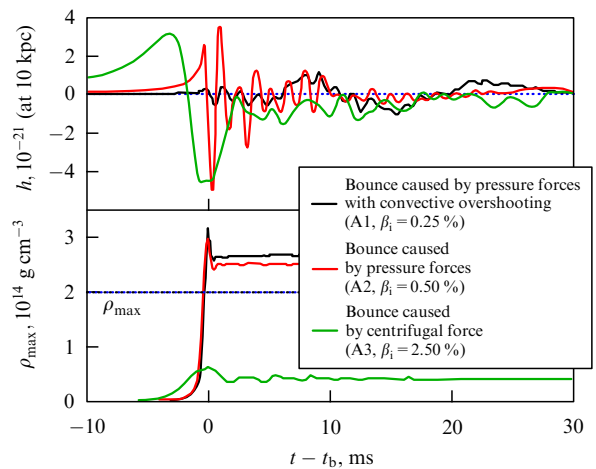

**Figure 4.** (Color online.) Gravitational wave amplitude $h$ and the maximum density $\rho_{\max}$ as functions of time for three core-collapse supernova models corresponding to different initial rotation parameters at a distance of 10 kpc [61]. A1, A2, A3 are the rotation profiles from [61] and $\beta_i = T/W$ is the initial rotational-to-gravitational energy ratio.

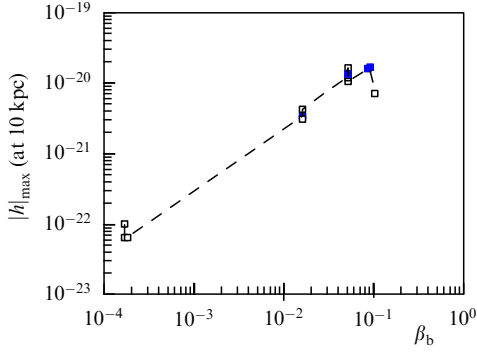

**Figure 5.** The maximum value of the relative GW amplitude $|h|_{\max, b}$ as a function of $\beta_b$ at a distance of 10 kpc [67]; $\beta_b = T/W$ is the rotational-to-gravitational energy ratio at the moment of maximum compression and bounce. For $\beta_b \leqslant 10\%$, $|h|_{\max, b}$ increases linearly with $\beta_b$. For $\beta_b > 10\%$, $|h|_{\max, b}$ starts decreasing due to the increase in the centrifuhal force.

(equal to the initial ratio of the rotational to gravitational energy). A catalog of GW waveforms for different parameters of the neutrino model is presented in [63]. The GW signal from magnetorotational supernovae was computed in [64–66] using two-dimensional, and in [67], three-dimensional, MHD models. We note that the effect of the magnetic field on the GW signal is significant only in very strong fields ($\sim 10^{12}$ G) of the pre-supernovae, when the magnetic field already affects the core collapse dynamics and the neutron star formation.

Figure 5 presents the maximum GW amplitude $h$ as a function of the rotation parameter $\beta_b$ computed in three-dimensional calculations [67]. Here, $\beta_b$ is the same as in (8) taken at the moment of maximum compression and bounce. Parameters of the GW signal from core-collapse supernovae for different explosion models are reviewed in [68, 69] (see also the references therein).

A rough general estimate of the GW power during nonspherical collapse was obtained in [56]. We estimate the GW energy from the collapse of a 1.2 $M_\odot$ core calculated in [51]. From formula (4), we obtain the energy of the GW pulse for the following parameters calculated at the bounce:

$$A_{\min} = 5.5 \times 10^5 \text{ cm}, \quad \frac{C_{\min}}{A_{\min}} = 0.50 \quad \text{for} \quad M = 0.24 M_\odot,$$
$$E_{GW} = 1.3 \times 10^{49} \text{ erg},$$

$$A_{\min} = 5.3 \times 10^6 \text{ cm}, \quad \frac{C_{\min}}{A_{\min}} = 0.79 \quad \text{for} \quad M = 0.3 M_\odot,$$
$$E_{GW} = 1.2 \times 10^{46} \text{ erg},$$

$$A_{\min} = 9.5 \times 10^6 \text{ cm}, \quad \frac{C_{\min}}{A_{\min}} = 0.83 \quad \text{for} \quad M = 0.5 M_\odot,$$
$$E_{GW} = 5.4 \times 10^{45} \text{ erg}, \qquad (9)$$

$$A_{\min} = 3 \times 10^7 \text{ cm}, \quad \frac{C_{\min}}{A_{\min}} = 0.89 \quad \text{for} \quad M = 0.8 M_\odot,$$
$$E_{GW} = 7.4 \times 10^{44} \text{ erg},$$

$$A_{\min} = 4.2 \times 10^7 \text{ cm}, \quad \frac{C_{\min}}{A_{\min}} = 0.94 \quad \text{for} \quad M = M_\odot,$$
$$E_{GW} = 5.9 \times 10^{44} \text{ erg}.$$

Here, $M$ is the mass of the star inside the isodense specified. The stellar collapse is significantly inhomogeneous. The energy of emitted GWs is approximately equal to the maximum value ($E_{GW} = 1.3 \times 10^{49}$ erg for mass $M = 0.24 M_\odot$) of energy in (9) for parameters corresponding to different masses $M$. This value is determined by the mass inside the isodense with the maximum degree of compression. The contribution from the outer layers of the collapsing star with a mass of 0.96 $M_\odot$ turns out to be insignificant. All the energy comes from the inner core corresponding to the maximum compressed isodense at the bounce. In the model in [51], the ratio of the rotational to gravitational energy at the bounce is $\beta_b \approx 0.028$. According to Fig. 5, this value is about one fourth that corresponding to the maximum emission of gravitational radiation energy. The characteristic bounce time is about $\Delta t \sim 0.7-0.8$ ms. The relative GW amplitude $h_{\alpha\beta}$ at a distance $r$ from the collapsing object is given by [17, 20]

$$h_{\phi\phi} = -h_{\theta\theta} = \frac{GM}{5rc^4}\sin^2\theta_0(\ddot{A}^2 - \ddot{C}^2), \quad h_{\theta\phi} = 0, \qquad (10)$$

where the angle $\theta_0$ is calculated relative to the $z$ axis of the spheroid. To estimate the maximum observation angle, the dimensionless GW amplitude at a distance $r$ from the source in Eqn (10) can be represented in the form

$$h_{\phi\phi} = -h_{\theta\theta} = \frac{GM}{5rc^4}\frac{A_{\min}^2}{(\Delta t)^2}. \qquad (11)$$

Using the values from (9), for the maximum GW flux corresponding to $M = 0.24 M_\odot$, we obtain the estimate

$$h = h_{\phi\phi} = -h_{\theta\theta} = \frac{6.7 \times 10^{-8}(0.48 \times 10^{33})}{5rc^4}$$
$$\times \frac{(5.5 \times 10^5)^2}{(0.75 \times 10^{-3})^2} \approx 1.4 \times 10^{-22}\frac{10 \text{ kpc}}{r}. \qquad (12)$$

Somewhat higher values of $h$ presented in Fig. 5 may be related to other parameters of the collapsing star, differences in the equation of state, treatment of the neutrino losses, or the use of different numerical schemes.

## 6. Formation of super-long gravitational waves from dark matter collapses

In [70], using Zel'dovich's model [71, 72], a phenomenological approach to the formation of dark matter structures was considered, in which the formation of super-long GWs was taken into account. The collapse of a rotating dark matter spheroid was examined. It was shown that for a realistic violent relaxation velocity [73] of a dust spheroid with mass $m$ and with the characteristic relaxation time $\tau_{\text{rel}}$ of the order of three Jeans times,

$$\tau_{\text{rel}} = 3\tau_J = 3\frac{2\pi}{\sqrt{4\pi G\rho}} = 6\pi\sqrt{\frac{a^2c}{3Gm}}, \qquad (13)$$

the oscillations decay, and after 10 periods their amplitude is $\sim 0.1$ times the initial value. Here, $a$ and $c$ denote the respective major and minor spheroid semi-axes. These oscillations and variable gravitational fields of the collapsing dark matter objects can leave imprints in observed cosmic microwave background (CMB) fluctuations. Super-long GWs, predominantly generated at the initial stage of the pancake formation, can also affect the CMB fluctuations and gravitational lensing of remote objects. To estimate the total

energy radiated during the collapse, formula (4) from [19] was used in [70, 74] in the form

$$E_{GW} \approx 0.01 \left(\frac{r_g}{A_{min}}\right)^{7/2} \left(\frac{A_{min}}{C_{min}}\right) Mc^2. \quad (14)$$

In the calculations in [70], the value

$$\frac{A_{min}}{C_{min}} \leqslant 100 \quad (15)$$

was reached, which significantly exceeded the applicability limits of formula (4) defined by relations (5); in the estimate below, we therefore use expression (2) for a nonrotating spheroid from [16].[1] The ratio $r_g/A_{eq}$ in the collapsing object (galaxy cluster) after reaching equilibrium can be estimated from the mean circular velocity of the individual galaxies, reaching the value $v_p \sim 2000$ km s$^{-1}$ [75],

$$\frac{r_g}{A_{eq}} \sim \left(\frac{v_g}{c}\right)^2 \approx \frac{4}{9} 10^{-4}. \quad (16)$$

The calculations in [70] suggest that the ratio of the equilibrium $A_{eq}$ to the minimum value $A_{min}$ is in the range from 2 to 5 depending on the parameters. For the mean value $A_{eq}/A_{min} = 3$, we obtain

$$\frac{r_g}{A_{min}} \approx \frac{4}{3} 10^{-4}. \quad (17)$$

Using (17) in (2), we obtain the mean energy $E_{GW}$ of GWs emitted during the collapse of a mass $M$:

$$E_{GW} \approx 10^{-15} Mc^2. \quad (18)$$

If all dark matter with a density $\rho_{DM}$ passes through the pancake formation stage, super-long GWs have the mean energy density of the Universe $\rho_{GW} \sim 10^{-15} \rho_{DM}$. By adopting the dark matter density [76]

$$\rho_{DM} = 3 \times 10^{-30} \text{ g cm}^{-3}, \quad (19)$$

we obtain the upper limit on the mean energy density of the super-long GWs in the Universe in the form

$$\mathcal{E}_{GW} \approx 10^{-15} \rho_{DM} c^2 \approx 3 \times 10^{-24} \text{ erg cm}^{-3}. \quad (20)$$

To estimate the GW amplitude, the following expressions can be used [17, 70]:

$$\mathcal{E}_{GW} = \frac{c^2}{16\pi G} \dot{h}^2, \quad \dot{h} = \omega h = \frac{2\pi c h}{\lambda}, \quad \lambda \sim 10 \text{ Mpc}. \quad (21)$$

As a result, the upper bound for the mean dimensionless metric perturbations $\bar{h}$ due to super-long GWs is

$$\bar{h} = \frac{2\lambda}{c^2}\left(\frac{G\mathcal{E}_{GW}}{\pi}\right)^{1/2} \approx 2 \times 10^{-11}. \quad (22)$$

The means of detection of such GWs are unclear as yet.

---

[1] In [16], GW radiation was considered only at the collapse stage. During the collapse of collisionless dark matter, an almost 'mirror' bounce occurs; therefore, the coefficient in formula (2) can be increased by a factor of two in subsequent estimates.

## 7. Conclusion

In nonspherical collapse, GWs are mostly radiated during the first bounce. The main frequency of the GW pulse during a core-collapse supernova explosion is about $10^3$ Hz.

The dimensionless amplitude $h$ of a GW emitted during a core-collapse supernova in our Galaxy at a distance of 10 kpc from Earth is sufficiently high, $\sim 10^{-22} - 10^{-20}$. Such a signal can be registered by the LIGO and Virgo detectors, especially after their upgrade, and in the case of maximum possible amplitude, by the running detectors TAMA 300 (Japan) [77], GEO 600 (UK–Germany) [78], and OGRAN (Optoacoustic GRavitational-wave ANtenna) (Russia), which is close to completion [79].

The research was supported by the Russian Science Foundation (project 15-12-30016).